\documentclass[aps,prl,twocolumn,superscriptaddress,longbibliography]{revtex4-2}
\usepackage{amsmath,amssymb}
\usepackage[pdftex]{hyperref,graphicx}
\hypersetup{colorlinks = true, urlcolor = blue, linkcolor = blue, citecolor = blue}
\usepackage{physics}
\usepackage{xcolor}
\usepackage{bm}
\usepackage{subfigure}
\usepackage{tikz}
\usetikzlibrary{positioning, fit, arrows.meta, calc, shapes.geometric}
\newcommand{\vtext}[2]{\rotatebox{90}{\parbox{#1}{\centering #2}}}

\newcommand{\boxfontsize}{\footnotesize}
\newcommand{\iofontsize}{\small}

\begin{document}
\title{Machine learning Majorana topology using unsupervised and supervised learning}
\author{Jacob R. Taylor}
\affiliation{Condensed Matter Theory Center and Joint Quantum Institute, Department of Physics, University of Maryland, College Park, Maryland 20742, USA}
\author{Haining Pan}
\affiliation{Department of Physics and Astronomy, Center for Materials Theory, Rutgers University, Piscataway, NJ 08854, USA}
\author{Sankar Das Sarma}
\affiliation{Condensed Matter Theory Center and Joint Quantum Institute, Department of Physics, University of Maryland, College Park, Maryland 20742, USA}

\begin{abstract}
In unsupervised learning, the training data for deep learning does not come with any labels, thus forcing the algorithm to discover hidden patterns in the data for discerning useful information. This, in principle, could be a powerful tool in identifying topological order since topology does not always manifest in obvious physical ways (e.g., topological superconductivity) for its decisive confirmation.  
The problem, however, is that unsupervised learning is a difficult challenge, necessitating huge computing resources, which may not always work.
In the current work, we combine unsupervised and supervised learning to establish that unlabeled (simulated) data in the Majorana splitting in realistic short disordered nanowires may enable not only a distinction between `topological' and `trivial', but also where their crossover happens in the relevant parameter space. This may be a useful tool in identifying topology in experimental Majorana nanowires.
\end{abstract}

\maketitle

\textit{Introduction.} Topological quantum computing (TQC) is a paradigm where appropriate manipulations of non-Abelian anyons, such as localized Majorana zero modes (MZM) in a topological superconductor (TSC), could lead to error-free fault-tolerant quantum computation \cite{kitaev2003faulttolerant,kitaev2001unpaired,freedman2002topological,dassarma2005topologically, nayak2008nonabelian,sarma2015majorana}.
Since the realistic theoretical predictions of the possible existence of MZMs in superconductor-semiconductor hybrid nanowire platforms in 2010 \cite{lutchyn2010majorana,sau2010generic,sau2010robustness, oreg2010helical},
a great deal of theoretical and experimental activities \cite{dassarma2023search,kouwenhoven2024perspective}
have focused on TQC in nanowires, with Microsoft dedicating a huge industrial effort on MZM in nanowires \cite{microsoftquantum2023inasal,aghaee2025interferometric,aghaee2025distinct,aghaee202620}.

In spite of this enormous effort since 2011 (including the huge recent Microsoft activities), a clear case for MZM is still elusive, mainly because of the inherent difficulties in identifying topological indicators in realistic systems in the presence of disorder (suppressing the TSC gap) and the relatively short wire length (thus, possibly creating MZM overlap suppressing their anyonic nature) \cite{liu2012zerobias,brouwer2011topological, bagrets2012class,akhmerov2011quantized, sau2013density,dassarma2021disorderinduced, zhang2021large, sarma2023density,pan2025majorana,sarma2023spectral,sau2025capacitancebased}.
A detailed topological gap protocol (TGP) has been introduced by Microsoft, but TGP only incorporates the necessary conditions for topology, and the sufficient conditions remain elusive \cite{pikulin2021protocol,pan2021threeterminal,rosdahl2018andreev}.
Developing decisive tools for identifying topological MZM in realistic systems remains the key open problem in TQC.

In this context, modern neural network-based Machine Learning (ML) techniques in the search for topological indicators in nanowires could be useful \cite{taylor2023machine,taylor2025vision,taylor2025mitigating}.
Although powerful from an abstract computer science viewpoint, one problem in these supervised ML techniques is that the training data sets must be labelled, identifying the individual topology for the training to lead to successful identification of MZM topology in the test data sets. Such supervised ML is relatively easy to do theoretically since every theoretical nanowire simulation used in training can be appropriately labeled as `topological' or `trivial' through explicit calculations of the topological visibility ($\text{TV}$)~\cite{dassarma2016how}.

Obviously, the experimental data, by definition, cannot come with any labels, and as such, doing supervised ML in the experimental context is a challenge as it would necessitate extensive theoretical simulations of the data to determine the appropriate label, thus partially negating the usefulness of the technique. What could be useful is an unsupervised or self-supervised ML algorithm that learns by itself how to label the data by discerning hidden patterns without the input training data carrying explicit topology labels. Of course, such unsupervised learning may not work, and also in the end, one must provide insight into the patterns discerned by unsupervised ML so that the patterns can have a one-to-one correspondence to MZM topological labels, i.e., `topological' or `trivial'. Unsupervised learning is also technically much more demanding than supervised ML.

In the current work, we introduce a comprehensive method to do unsupervised ML (UML) for MZM topology identification, by first using UML to identify underlying patterns in the nanowire data, and then utilizing supervised ML (SML) as an input to label the topology.
The combining of UML/SML in a seamless manner playing complementary roles may potentially solve the conundrum of identifying MZM topology in realistic systems.
We explicitly check the fidelity of the method on test data, finding excellent fidelity, establishing this as a possible breakthrough technique in MZM TQC. Furthermore, the specific supervised learning model we use here provides a new method for predicting the topological scattering invariant that is more robust and less parameter-specific than the previous transport-based machine learning methods~\cite{taylor2025vision}.
We note that the current work completely sidesteps the problematic issue of zero bias conductance peaks in Majorana nanowires~\cite{dassarma2023search}, and focuses directly on the central role of MZM splitting which is the sole determinant of topology~\cite{kitaev2001unpaired}.

The specific quantity used in our UML/SML is the MZM splitting which must be exponentially small~\cite{kitaev2001unpaired} in the topological regime so that the MZM are actual zero energy anyonic Majorana bound states~\cite{cheng2009splitting,brouwer2011probability}, and not just low-energy trivial fermionic Andreev bound states which occur generically in the nanowire platform, hugely complicating the MZM identification~\cite{pan2020physical,pan2020generic}.
Therefore, whether the MZM splitting is exponentially small or not is a direct measure of the topology or not in nanowires. We use MZM splitting with no labels for our UML, finding, rather amazingly, precisely 2 or 3 clusters coming out of UML. The 2-clusters happen when the wire length is large [compared with superconducting (SC) coherence length] and the disorder is small (compared with SC gap), indicating the explicit presence only of topological or trivial phases (depending on the system parameters). By contrast, 3-clusters emerge as the UML-discerned hidden pattern only when the wire length is not necessarily large and disorder is not necessarily small, and the UML then finds three distinct patterns in the unlabeled data: topological and trivial as well as an intermediate regime in between where the topology is ill-defined and may be dominated by ABS contamination. (Much of the current experimental data seems to be in the intermediate regime.) The fidelity of the UML drops precipitously if we force the algorithm to find more than three patterns in the data, clearly establishing that UML has been effective (with no prompting/labeling) in identifying the correct topological patterns occurring in realistic samples.
We emphasize that the UML algorithm is entirely physics-agnostic: it encodes no knowledge of the Bogoliubov--de~Gennes (BdG) Hamiltonian, the exponential dependence of MZM splitting on wire parameters in the topological regime, the topological quantum phase transition (TQPT), or even the distinction between topological and trivial phases.
That such a physics-ignorant algorithm discovers precisely 2 or 3 physically meaningful clusters---rather than some arbitrary number---purely from the structure of the unlabeled data is a remarkable and unanticipated finding, enabled by the autoencoder's capacity to learn efficient compressed representations of the input.
Supervised learning enters only at the final stage to assign physical labels (topological, trivial, or intermediate) to the clusters that UML has already identified; the number and the internal structure of the clusters are determined entirely by the unsupervised algorithm.

\textit{Model and data.} We start with the standard semiconductor-superconductor single-band model
\begin{equation}\label{eq:hatH}
    \hat{H}=\frac{1}{2}\int dx \Psi^\dagger(x) \left( H_{\text{SM}}+H_{\text{Z}}+H_{\text{SC}}+H_{\text{dis}} \right) \Psi(x),
\end{equation}
where $\Psi(x)=(\psi_\uparrow(x),\psi_\downarrow(x),\psi_\downarrow^\dagger(x),-\psi_\uparrow^\dagger(x))^\intercal$ is the Nambu spinor defined such that the single-particle Hamiltonian of the SM part $H_{\text{SM}}$, Zeeman field part $H_{\text{Z}}$, proximity-induced superconducting part $H_{\text{SC}}$, and disorder part $H_{\text{dis}}$ are given by~\cite{lutchyn2010majorana,sau2010generic,sau2010robustness, oreg2010helical,dassarma2023search}:
\begin{equation}\label{eq:H}
    \begin{split}
        H_{\text{SM}}&=\left( -\frac{\partial_x^2}{2m^*}-i\alpha \partial_x\sigma_y-\mu \right)\tau_z, ~  H_{\text{Z}}=V_z \sigma_x,\\
        H_{\text{SC}}&=\Delta \tau_x, ~ H_{\text{dis}}=V(x)\tau_z 
    \end{split}
\end{equation}
where the definitions and the values of the parameters are shown in Table~\ref{tab:table1}. 
We are using realistic nanowire parameters, and zero temperature is justified by the low ($\sim~$  20 mK) experimental temperatures along with the fact that the TSC gap is $\sim$ 400mK or more.~\cite{microsoftquantum2023inasal,aghaee2025interferometric,aghaee2025distinct} 

\begin{figure*}[htbp]
    \centering
    \begin{minipage}[c]{0.2\textwidth}
        \centering
        \includegraphics[width=\linewidth]{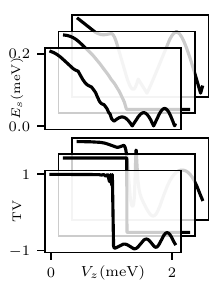}
    \end{minipage}%
    \hfill
    \begin{minipage}[c]{0.6\textwidth}
        \centering
        \resizebox{\linewidth}{!}{%
        \begin{tikzpicture}[
    boxfont/.style={font=\normalfont\footnotesize},
    labelfont/.style={font=\normalfont\bfseries\footnotesize},
    >=Stealth,
    node distance=0.6cm and 0.3cm,
    layer/.style={rectangle, draw=blue!60!black, fill=blue!10, thick,
        minimum width=0.6cm, minimum height=3.0cm, align=center,
        rounded corners=2pt, inner sep=2pt, boxfont},
    layer_med/.style={layer, minimum height=2.5cm},
    layer_small/.style={layer, minimum height=1.9cm},
    bottleneck/.style={rectangle, draw=black!70, fill=yellow!10, thick,
        minimum width=0.8cm, minimum height=2.7cm, align=center,
        rounded corners=5pt, inner sep=2pt, font=\normalfont\bfseries\footnotesize},
    container/.style={draw=blue!20, thick, inner xsep=0.2cm, inner ysep=0.4cm, rounded corners=2pt,
        label={[anchor=north, labelfont, text=black!70, align=center]north:#1}},
    arrow/.style={->, thick, draw=blue!60!black}
]

    \node (enc1) [layer] {\vtext{2.8cm}{1D Conv}};
    \node (enc2) [layer_med, right=0.3cm of enc1] {\vtext{2.2cm}{1D Conv}};
    \node (enc3) [layer_small, right=0.3cm of enc2] {\vtext{1.7cm}{1D Conv}};
    \node (enc4) [layer_small, right=0.3cm of enc3, minimum width=0.7cm] {\vtext{1.7cm}{Linear}};
    \node (encoder_group) [container={Encoder}, fit=(enc1)(enc4), minimum height=4.2cm] {};

    \node (latent) [bottleneck, right=0.3cm of encoder_group] {\vtext{2.5cm}{15-dim neurons}};
    \node (latent_group) [container={Latent\\Space}, fit=(latent), minimum height=4.2cm] {};

    \node (dec1) [layer_small, right=0.3cm of latent_group, minimum width=0.7cm] {\vtext{1.7cm}{Linear}};
    \node (dec2) [layer_small, right=0.3cm of dec1] {\vtext{1.7cm}{1D Conv\textsuperscript{T}}};
    \node (dec3) [layer_med, right=0.3cm of dec2] {\vtext{2.2cm}{1D Conv\textsuperscript{T}}};
    \node (dec4) [layer, right=0.3cm of dec3] {\vtext{2.8cm}{1D Conv\textsuperscript{T}}};
    \node (decoder_group) [container={Decoder}, fit=(dec1)(dec4), minimum height=4.2cm] {};

    \draw[arrow] ($(enc1.west)+(-1.5cm,0)$) -- node[midway, above, boxfont] {Compress} (enc1.west);
    \draw[arrow] (dec4.east) -- node[midway, above, boxfont] {Reconstruct} ($(dec4.east)+(1.5cm,0)$);

    \draw[arrow] (enc1)--(enc2)--(enc3)--(enc4)--(latent)--(dec1)--(dec2)--(dec3)--(dec4);

\end{tikzpicture}%
        }%
    \end{minipage}%
    \hfill
    \begin{minipage}[c]{0.2\textwidth}
        \centering
        \includegraphics[width=\linewidth]{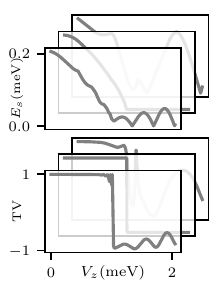}
    \end{minipage}
    \caption{
        Autoencoder architecture used for dimensionality reduction in the unsupervised learning framework. The network consists of a 1D convolutional encoder (with stride size of 2) and decoder with a symmetric architecture in reverse order. Left column shows examples of input data: Majorana energy splittings $E_s(V_z)$ and topological visibility $\text{TV}(V_z)$. The encoder compresses these into a 15-dimensional latent vector, which the decoder reconstructs back to the original input format. Training minimizes the mean-squared error between input and reconstructed output.
   }
    \label{fig:fig1}
\end{figure*}

Our dataset consists of two ingredients: the MZM energy splitting $E_s(V_z)$~\cite{kitaev2001unpaired,cheng2009splitting} and the topological visibility $\text{TV}(V_z)$~\cite{dassarma2016how}, both as a function of the Zeeman field $V_z$. 
The Majorana energy splitting $E_s$ is defined as the energy of the lowest-lying state in the nanowire, which is obtained by numerically discretizing Hamiltonian \eqref{eq:hatH} with a fictitious lattice constant $a=10~\text{nm}$ and then diagonalizing the resulting tight-binding model. We emphasize that this splitting does not assume the existence of MZMs; it is agnostic to the source, whether topological or trivial.
(The lattice constant `$a$' is essentially a measure of the disorder correlation length in the system.)
The topological visibility ($\text{TV}$) is defined as the determinant of the reflection matrix at zero energy~\cite{dassarma2016how}, computed using the KWANT package \cite{groth2014kwant}, which takes values between $-1$ (topological) and $+1$ (trivial), and zero at the topological quantum phase transition point.
Examples of both $E_s(V_z)$ and $\text{TV}(V_z)$ are shown in the left column in Fig.~\ref{fig:fig1}.
For each combination of $(L, \sigma)$ in Table~\ref{tab:table1}, we generate 500 independent disorder realizations, each producing a pair of $E_s(V_z)$ and $\text{TV}(V_z)$ curves. The same dataset is used for both the unsupervised and supervised learning analyses.
We emphasize a crucial point: $\text{TV}$ is continuous between $\pm 1$ and is not binary, as it would become for an infinite system with no disorder~\cite{kitaev2001unpaired}--- the topological quantum phase transition is still defined by the vanishing of $\text{TV}$~\cite{dassarma2016how}. 
This means that $\text{TV}(V_z)$ for a given disorder realization can indicate topology at some $V_z$ values and triviality at others, so we are asking whether given disorder strength will lead to the existence of any topological phase.

\begin{table}[t]
\centering
\caption{Parameters in the Majorana nanowire simulations adapted from \cite{lutchyn2018majorana}.}
\label{tab:table1}
\begin{tabular}{c c}
\hline\hline
effective mass $m^{*}$ & $0.01519~m_{e}$ \\
chemical potential $\mu$ & $1~\mathrm{meV}$ \\
spin-orbit coupling $\alpha$  & $0.5~\mathrm{eV \AA}$ \\
constant SC gap $\Delta$ & $0.2~\mathrm{meV}$ \\
wire length $L$  & $0.6$--$15~\mu\mathrm{m}$ \\
Gaussian disorder strength $\sigma$ & $0.1\text{--}6~\mathrm{meV}$ \\
temperature & $0$ \\
dissipation & $0$ \\
SC coherence length $\xi$  
& $0.78~\mu\mathrm{m}$ \\
localization length $\ell_{\mathrm{loc}} = \dfrac{v_{F}^{2}}{\sigma^{2} a}$ 
& $2.31~\mu\mathrm{m}~ \frac{1~\mathrm{meV}^2}{\sigma^{2}}$ \\
\hline\hline
\end{tabular}

\end{table}

\begin{figure}[htbp]
    \centering
    \includegraphics[width=3.4in]{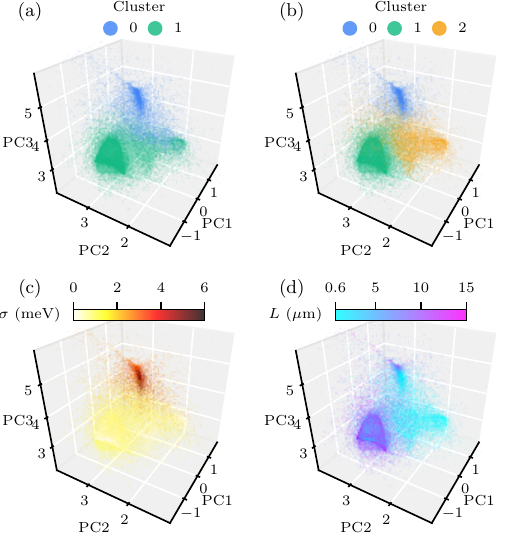}
    \caption{Top row: Latent space visualization from the auto-encoder principal component analysis (PCA), displaying the first three principal components (PC1, PC2, and PC3) for (a) $k=2$ and (b) $k=3$ cluster classifications obtained from $k$-means clustering. Each data point represents one disorder realization, compressed from the complete $E_s(V_z)$ and $\text{TV}(V_z)$ curves via the autoencoder in Fig.~\ref{fig:fig1}. Bottom row: Physical parameters associated with each latent space point: (c) disorder strength $\sigma$ and (d) wire length $L$.}
    \label{fig:fig2}

\end{figure}

\textit{Unsupervised learning.}
We first unveil hidden patterns in the unlabeled Majorana data using UML, which contains two main steps: (i) dimensionality reduction using an autoencoder neural network, and (ii) clustering in the latent space using the $k$-means algorithm.
The autoencoder architecture is shown in Fig.~\ref{fig:fig1}, where we employ 1D convolutional layers to compress the input data of $E_s(V_z)$ and $\text{TV}(V_z)$ into a 15-dimensional latent space, and then reconstruct the original input from the latent vector. (See Sec.~\ref{sec:nnarchitecture} in the Appendix ~\cite{supplemental} for details of the autoencoder architecture.)
Here, the middle latent space is expected to capture the most important features of the input data.

At this point, we effectively compress the original high-dimensional data into a low-dimensional latent space, where we can then perform clustering using the $k$-means algorithm (see Sec.~\ref{sec:kmeans} in the Appendix for details of the $k$-means algorithm).
Figure~\ref{fig:fig2} shows the clustering results in the latent space for (a) two-cluster and (b) three-cluster classifications, visualized using principal component analysis (PCA) or (more precisely the latent space vector) to show the first three principal components (PC1, PC2, and PC3).

To understand the physical meaning of the clusters identified in the latent space by the $k$-means algorithm, we color each data point according to its associated disorder strength $\sigma$ [Fig.~\ref{fig:fig2}(c)] and system size $L$ [Fig.~\ref{fig:fig2}(d)].  
We find that `Cluster 0' (blue) in both two- and three-cluster classifications corresponds to the strong disorder regime (large $\sigma$), while `Cluster 1' (green) corresponds to the weak disorder regime (small $\sigma$).

\begin{figure*}[htbp]
    \centering
    \includegraphics[width=6.8in]{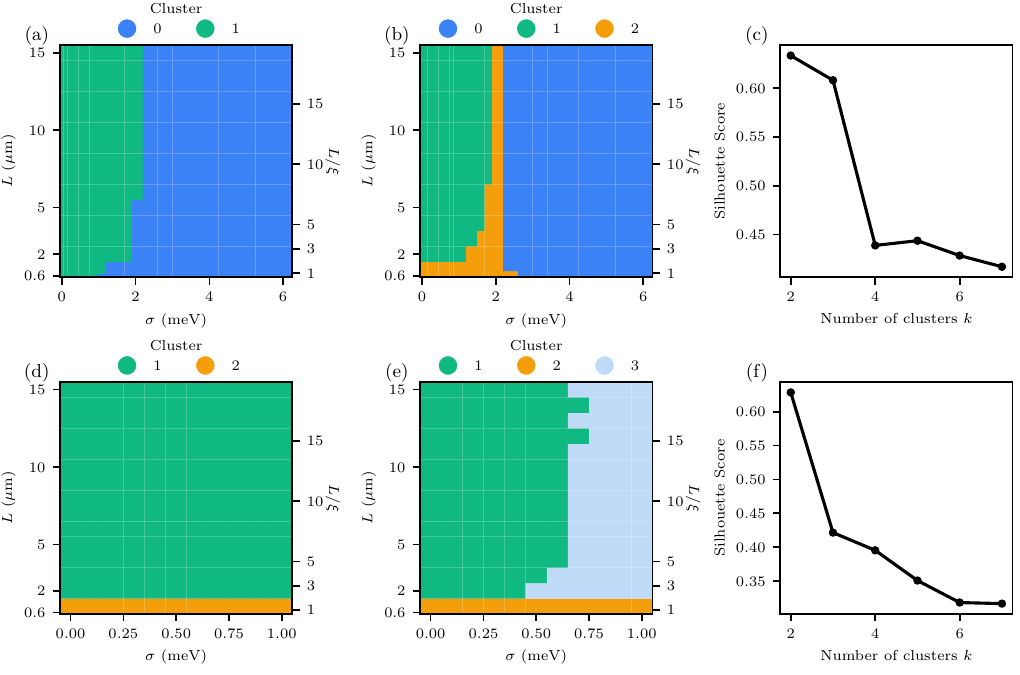}
    \caption{
    Top row: Unsupervised phase diagrams in the disorder-length ($\sigma$-$L$) parameter space, obtained by mapping latent space clusters back to physical parameters.
    (a) Two-cluster classification reveals a phase boundary between weak-disorder (Cluster 1, green) and strong-disorder (Cluster 0, blue) regimes.
    (b) Three-cluster classification identifies an additional intermediate regime (Cluster 2, orange) at the crossover.
    Each point is labeled by the majority cluster among 100 disorder realizations.
    (c) Silhouette score as a function of cluster number $k$, showing that $k=2$ and $k=3$ have comparable quality, but the score drops sharply beyond $k=3$.
    Bottom row (d-f): Same clustering analysis for restricted disorder range $\sigma \in [0,1]$ meV, where (f) indicates optimal clustering at $k=2$.
    }
    \label{fig:fig3}
\end{figure*}

This heuristic observation allows us to map the clustering results in the latent space back to the original physical parameter space spanned by $\sigma$ and $L$, and thus obtain unsupervised phase diagrams as shown in the top row of Fig.~\ref{fig:fig3}.
Here, without any \textit{a priori} knowledge, we find that the clusters in the latent space naturally correspond to distinct phases in the physical parameter space.
For the two-cluster classification [Fig.~\ref{fig:fig3}(a)], we find a phase boundary separating the weak-disorder topological phase (Cluster 1 in green) and the strong-disorder trivial phase (Cluster 0 in blue). 
Here, each data point in the parameter space is assigned a label according to the majority cluster among its 100 disorder realizations.
This ensemble perspective is essential in disordered one-dimensional superconductors, where Griffiths rare-region physics and the associated infinite-randomness behavior can generate extremely broad sample-to-sample distributions and atypical low-energy features in individual disorder realizations~\cite{motrunich2001griffiths}. Thus, topology assessment based on a single realization, or on a small number of hand-selected features, can be misleading. The autoencoder-based approach instead learns correlations across the full distribution of $E_s(V_z)$ and $\text{TV}(V_z)$ curves, whose high-dimensional structure is difficult to label reliably by hand.
The phase boundary (crossover) shifts to larger disorder strength for longer wire lengths, consistent with the expectation that longer wires are more robust against disorder.
For the three-cluster classification [Fig.~\ref{fig:fig3}(b)], we find an additional cluster (Cluster 2 in orange) that corresponds to an intermediate crossover disorder regime.

\textit{Stability of clustering.} We find that the clustering results are robust against different runs of the dimensionality reduction in the autoencoder, as well as different initializations of the $k$-means algorithm (see Sec.~\ref{sec:kmeans} in the Appendix for details).
We also perform a heuristic elbow-method analysis to determine the optimal number of clusters $k$ by evaluating a standard clustering-quality metric, the Silhouette score (see Sec.~\ref{sec:silhouette} in the Appendix for details), in Fig.~\ref{fig:fig3}(c).
Here, we find that $k=2$ and $k=3$ yield comparable separation quality, but the score drops sharply from $k=3$ to $k=4$, indicating that having more than three clusters does not improve the clustering quality.

We further restrict the disorder range to $\sigma \in [0,1]$ meV (which is the estimated low-disorder range in the current MSFT experiments), and repeat the $k$-means clustering in the latent space, obtaining the phase diagrams shown in the bottom row of Fig.~\ref{fig:fig3}.
Here, we find that the two-cluster classification in Fig.~\ref{fig:fig3}(d) is consistent with the phase diagram in Fig.~\ref{fig:fig3}(b) within its zoomed-in disorder range.
The three-cluster classification in Fig.~\ref{fig:fig3}(e) reveals an additional intermediate cluster which is a clear artifact from the Silhouette score analysis in Fig.~\ref{fig:fig3}(f), indicating that the clustering quality is best for $k=2$ in the low-disorder regime.

\begin{figure}[htbp]
    \centering
    \includegraphics[width=1\linewidth]{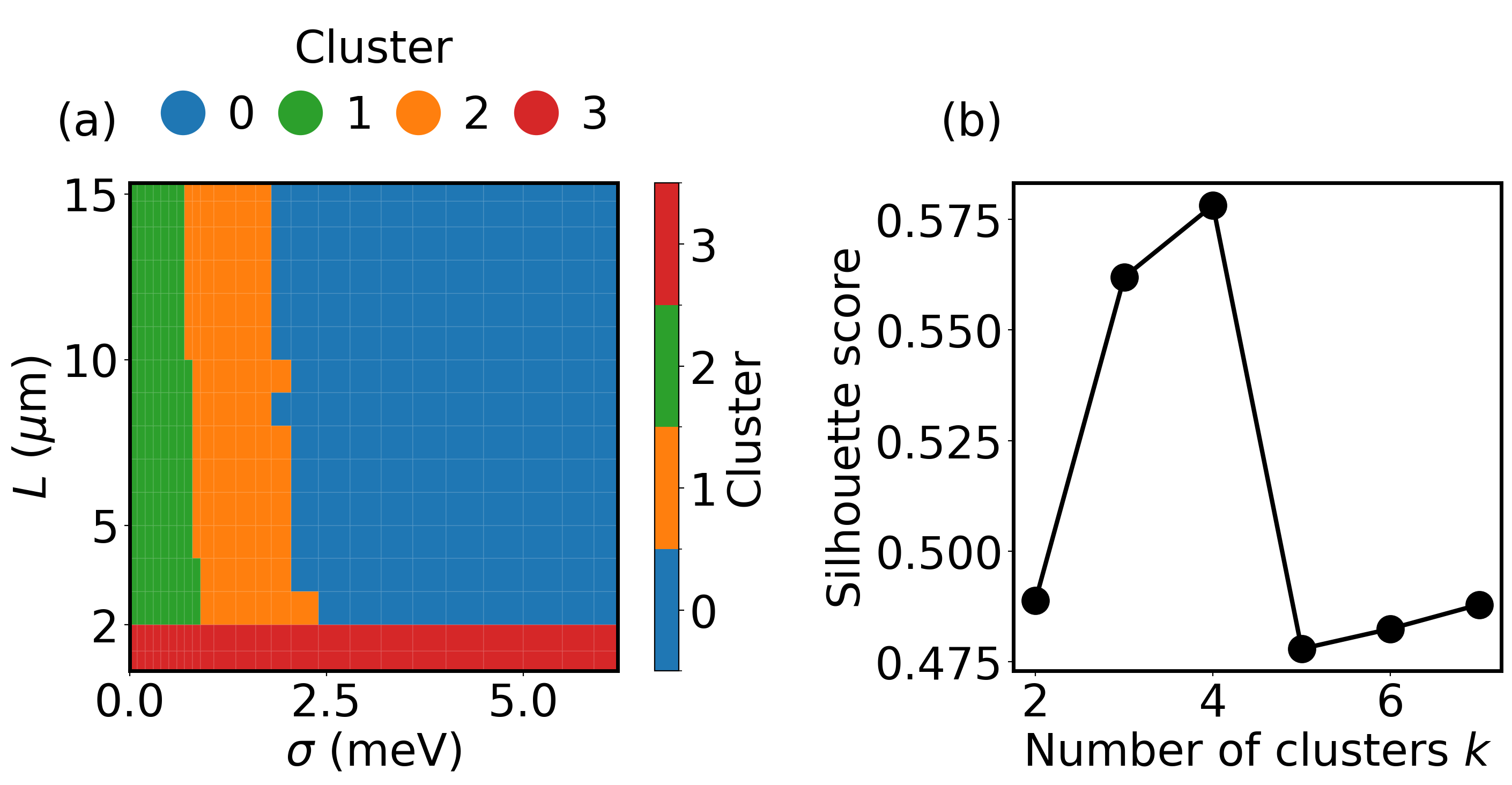}
    \caption{Unsupervised phase diagrams in disorder-length ($\sigma$-$L$) space, obtained by mapping latent-space clusters back to physical parameters using $E_s$ and $L$ as inputs (but only rewarding $E_s$ reconstruction).
    (a) Four-cluster classification separates weak-disorder, strong-disorder, intermediate-disorder, and short-length regimes; each point is labeled by the majority cluster over 100 disorder realizations.
    (b) Silhouette score versus cluster number $k$, showing optimal clustering at $k=4$ with a drop for larger $k$.}
    \label{fig:figUSLNOTV}
\end{figure}

For comparison, we also consider an $E_s$-only variant in which $E_s(V_z)$ and $L$ are used as inputs, but the autoencoder is rewarded only for reconstructing $E_s(V_z)$. This modified scheme gives phase diagrams roughly consistent with the TV-inclusive analysis, as shown in Fig.~\ref{fig:figUSLNOTV}, but the clustering is less stable because the latent space tends to organize by length; see Appendix Sec.~\ref{sec:onlyes} for more details.

\begin{figure*}[htbp]
    \centering
    \begin{subfigure}
        \centering
\begin{tikzpicture}[
    font=\normalfont,
    >=Stealth,
    node distance=0.8cm and 1.0cm,
    layer/.style={
        rectangle,
        draw=blue!60!black,
        fill=blue!10,
        thick,
        minimum width=3.5cm,
        minimum height=0.8cm,
        align=center,
        rounded corners=2pt,
        font=\normalfont\boxfontsize
    },
    small_layer/.style={
        layer,
        minimum width=2.5cm,
        fill=blue!5
    },
    process/.style={
        circle,
        draw=black!70,
        thick,
        minimum size=0.6cm,
        fill=white
    },
    container/.style={
        draw=black!40,
        dashed,
        inner sep=0.3cm,
        rounded corners=5pt,
        label={[anchor=south east, font=\footnotesize\bfseries, text=black!60]south east:$3\times$ }
    },
    conn/.style={->, thick},
    concat_node/.style={
        circle,
        draw=black,
        fill=white,
        inner sep=1pt,
        minimum size=0.4cm,
        path picture={
            \draw[black] (path picture bounding box.north) -- (path picture bounding box.south);
            \draw[black] (path picture bounding box.west) -- (path picture bounding box.east);
        }
    }
]

    \node (input_es) [font=\iofontsize] {$E_s(V_z)$};

    \node (enc_conv) [layer, below=0.6cm of input_es] {1D Conv Layer};
    \node (enc_pool) [layer, below=0.3cm of enc_conv] {Adaptive Average Pooling};

    \node (encoder_group) [container, fit=(enc_conv) (enc_pool)] {};

    \node (sum) [process, below=0.7cm of encoder_group, label={[label distance=2pt, align=center]right:\footnotesize element-wise\\addition}] {$+$};

    \node (dec_interp) [layer, below=0.5cm of sum] {Interpolative Expansion};
    \node (dec_conv) [layer, below=0.3cm of dec_interp] {1D Conv Layer};
    \node (dec_film) [layer, below=0.3cm of dec_conv] {FiLM Layer};

    \node (decoder_group) [container, fit=(dec_interp) (dec_conv) (dec_film)] {};

    \node (panel_a) [left=3.5cm of input_es, font=\iofontsize, yshift=-0.2cm] {(a)};

    \node (input_L) [right=0.3cm of panel_a, font=\iofontsize] {$L$};

    \node (concat) [concat_node, below=2.6cm of input_L, label={[label distance=2pt]left:\footnotesize concat}] {};

    \node (dense_shared) [small_layer, below=0.3cm of concat] {Dense};

    \node (linear) [small_layer, minimum width=1cm, right=0.4cm of dense_shared] {Linear};

    \node (dense_sigma) [small_layer, below=2.2cm of dense_shared] {Dense};
    \node (disorder_out) [small_layer, below=.9cm of dense_sigma] {Disorder Output};
    \node (output_sigma) [below=0.4cm of disorder_out, font=\iofontsize] {$\sigma$};

    \node (out_conv) [layer] at (decoder_group |- disorder_out) {1D Conv (to scalar) $\text{TV}$ Output};
    \node (output_tv) [below=0.4cm of out_conv, font=\iofontsize] {$\text{TV}(V_z)$};

    \draw[conn]
        (input_es) -- (enc_conv)
        (enc_conv) -- (enc_pool)
        (enc_pool) -- (sum);

    \draw[conn]
        (input_L) -- (concat)
        (concat) -- (dense_shared)
        ($(enc_pool.south)!0.5!(sum.north)$) |- (concat.east);

    \draw[conn]
        (dense_shared) -- (linear)
        (linear) -- (sum);

    \draw[conn] (dense_shared) -- (dense_sigma)
        node[midway, left, rotate=90, anchor=south, font=\footnotesize, align=center] {Global Context\\Vector};
    \draw[conn]
        (dense_sigma) -- (disorder_out)
        (disorder_out) -- (output_sigma);

    \draw[conn]
        (sum) -- (dec_interp)
        (dec_interp) -- (dec_conv)
        (dec_conv) -- (dec_film)
        (dec_film) -- (out_conv)
        (out_conv) -- (output_tv);

    \draw[conn] (dense_shared.south) -- ++(0,-0.8cm) -| ($(dec_film.west) + (-0.5cm,0)$) -- (dec_film.west);

\end{tikzpicture}
    \end{subfigure}
    \hfill
    \begin{subfigure}
        \centering
        \includegraphics[width=4.1in]{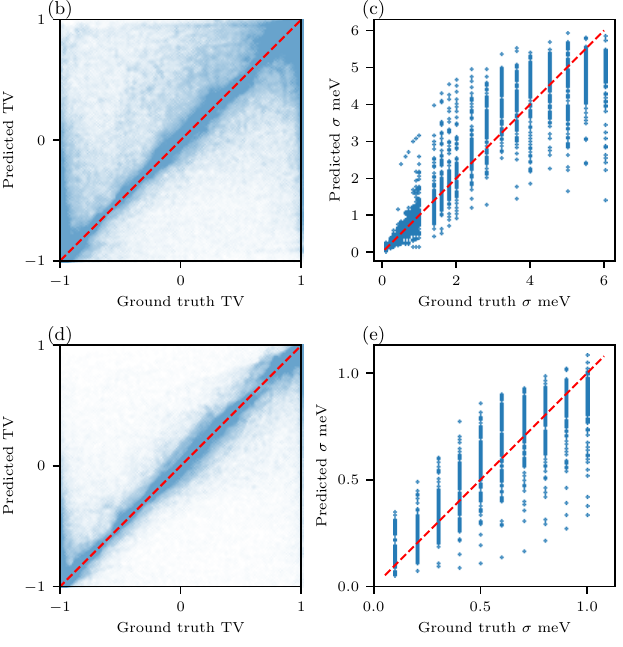}
    \end{subfigure}
    \caption{(a) Architecture of the supervised learning neural network based on 1D convolutional encoder-decoder. Inputs are experimentally measurable quantities ($E_s(V_z)$ and wire length $L$); outputs are theoretically accessible quantities (topological visibility $\text{TV}(V_z)$ and disorder strength $\sigma$) that are not directly measurable in experiments.
   (b) Predicted versus ground-truth topological visibility for the full disorder range $\sigma \in [0,6]$ meV, showing good correlation along the diagonal (red dashed line indicates perfect prediction).
   (c) Predicted versus ground-truth disorder strength for $\sigma \in [0,6]$ meV; accuracy is higher at weak disorder and decreases at strong disorder.
   (d, e) Same predictions for the restricted weak-disorder range $\sigma \in [0,1]$ meV, demonstrating significantly improved accuracy for both $\text{TV}$ and $\sigma$.
   }
    \label{fig:fig4}
\end{figure*}

\textit{Supervised learning.} In the previous unsupervised learning analysis, we have already identified the internal patterns in the unlabeled Majorana data, which naturally correspond to distinct phases induced by disorder, but we do not know the topology associated with the distinct phases since UML only discerns patterns, and cannot provide labels for the patterns.
 
However, we find that the most stable version of this unsupervised learning scheme uses both $E_s(V_z)$ and the topological visibility $\text{TV}(V_z)$ as inputs. A naive experimentally accessible variant based on $E_s(V_z)$ and $L$ tends to cluster by length rather than topology, although the modified scheme in Sec.~\ref{sec:onlyes} of the Appendix can still recover approximate phase structure.
This motivates a supervised learning framework to predict the topological visibility $\text{TV}(V_z)$, which is accessible in theory, from experimentally measurable inputs.
Therefore, in actual experiments, the supervised learning model can be the upstream component before the unsupervised learning (for more details, the semi-supervised experimentally compatible process can be found in Sec.~\ref{sec:semisupervised} of the Appendix). 
This supervised component is distinct from topological-gap-protocol-type binary criteria, which ask whether a negative scattering invariant occurs somewhere in a parameter region and therefore require threshold choices. By contrast, our network predicts the continuous scattering invariant over the full Zeeman-field range from the Majorana splitting alone, avoiding hand-tuned visual thresholds and using a reduced input compared with full conductance-based ML approaches. For experimental application, it is necessary that a theoretical model exists which captures the physics of the device, even if its parameters are not known. The main possible failure mode is therefore a mismatch between the theoretical training ensemble and the actual device, which can be mitigated in practice by sampling uncertain device parameters broadly and checking consistency across the relevant parameter subspace. Of course, for the analysis of the experimental data using our proposed technique, an extended model with more parameters may be necessary, which may necessitate larger computational resources in carrying out the machine learning, but this is not an essential problem since our work has the key ingredients of short wires and disorder inherently built in at the Hamiltonian level. This restricted input limits model-specific fine tuning, while robustness to measurement error can be incorporated and tested directly through noise augmentation of the training data.

To this end, we design a supervised learning neural network (see Fig.~\ref{fig:fig4}(a)) based on a 1D convolutional encoder-decoder architecture, where the inputs are two experimentally measurable quantities: the Majorana energy splitting $E_s(V_z)$ and the wire length $L$, while the outputs are the topological visibility $\text{TV}(V_z)$ and the disorder strength $\sigma$, which are not directly measurable in experiments.
We split the dataset into a training 95\% and a test on 5\% withheld data.
We find generally good prediction accuracy with $R^2$ of $0.861$ and $0.922$ errors of $\pm 0.654$ and $\pm 0.242$ for $\sigma(V_z)$ and $\text{TV}(V_z)$ respectively.
In Fig.~\ref{fig:fig4}(b), we compare the predicted topological visibility (vertical axis) with the ground-truth values (horizontal axis) for the full disorder range $\sigma \in [0,6]$ meV, finding a good correlation along the diagonal line (red dashed line).
In Fig.~\ref{fig:fig4}(c), the predicted disorder strength (vertical axis) is compared with the ground-truth values (horizontal axis) for the full disorder range $\sigma \in [0,6]$ meV, where we find that the model performs well at weak disorder but its accuracy decreases at strong disorder. (This is not a problem for experiments, which must avoid the very strong disorder regime anyway.)

The prediction accuracy improves significantly when we restrict the disorder range to $\sigma \in [0,1]$ meV, as shown in Figs.~\ref{fig:fig4}(d) and (e) for the topological visibility and disorder strength, respectively, where we find a much better correlation along the diagonal line and report higher accuracy with $R^2$ of $0.722$ and $0.971$ errors of $\pm 0.152$ and $\pm 0.157$ for $\sigma(V_z)$ and $\text{TV}(V_z)$ respectively. For the detailed accuracy profile as a function of disorder strength and wire length, see Fig.~\ref{fig:accuracy_weak} and Fig.~\ref{fig:accuracy_strong}, respectively, for weak and strong disorder in the Appendix. We further demonstrate that the supervised prediction of $\text{TV}(V_z)$ is robust to measurement error in $E_s(V_z)$, maintaining accuracy even at high noise levels (see Fig.~\ref{fig:figRobustness} in the Appendix).

\textit{Discussion.} One question remains open which is manifestly beyond the scope of the current work: Will our proposed USL + SML techniques work on the actual experimental data in discerning the underlying topology in experimental nanowires as successfully as demonstrated in the current work for the simulated minimal nanowire model?  This is an unanswerable question for the future, which can only be answered after the experimental data are used for this purpose. All we can assert is that we see no obvious reason for the technique not to work if sufficient experimental training data are available. We emphasize for the TV free use of USL, all one needs are the experimental data without any simulations since the whole point of the USL is that no labeling is needed.  If the experimental data clearly shows, for either approach, the correct number of clusters that will be a very big deal already since it would establish that the experiments are indeed using the right type of samples which have underlying patterns corresponding to only two possibilities with a boundary separating them.  We mention the obvious that we cannot guarantee that the technique will surely succeed with experimental data, there are never such guarantees in physics, but we see no obvious reason why it would not.

\textit{Conclusion.} We have carried out an unsupervised machine learning of unlabeled Majorana mode splitting in the current experimentally active TQC nanowire platform, finding to our surprise that the splitting naturally falls into machine-identified 2- and 3-clusters  respectively for weak and strong disorder, where we can identify the clusters through supervised learning input to be topological or trivial in the 2-clusters, with the 3-clusters for higher disorder also containing an intermediate phase of ill-defined topology. We also provide an alternative scheme that roughly replicates these phases without supervised learning which can be directly applied to experiment.  Our work should be useful in the classification of Majorana experimental data, particularly in the context of ruling out the parameter regimes where topology may not manifest in realistic nanowires. 

\textit{Acknowledgements.} 
This work is supported by the Laboratory for Physical Sciences through the Condensed Matter Theory Center at Maryland. 
JRT thanks the Joint Quantum Institute for additional support. 
HP is supported by US-ONR grant No. N00014-23-1-2357.
\bibliography{Paper_Splitting_ML}
\appendix
\clearpage
\renewcommand{\thesection}{\Roman{section}}
\vspace{3cm}
\onecolumngrid
\begin{center}
    {\bf \large Appendix for ``Machine learning Majorana topology using unsupervised and supervised learning''}
\vspace{1cm}
\end{center}
\twocolumngrid
\setcounter{page}{1}
\setcounter{secnumdepth}{3}
\setcounter{equation}{0}
\setcounter{figure}{0}
\renewcommand{\theequation}{S-\thesection.\arabic{equation}}

\renewcommand{\thefigure}{A\arabic{figure}}
\renewcommand\figurename{Supplementary Figure}
\renewcommand\tablename{Supplementary Table}

\begin{figure*}[ht]
    \centering \includegraphics[width=\textwidth]{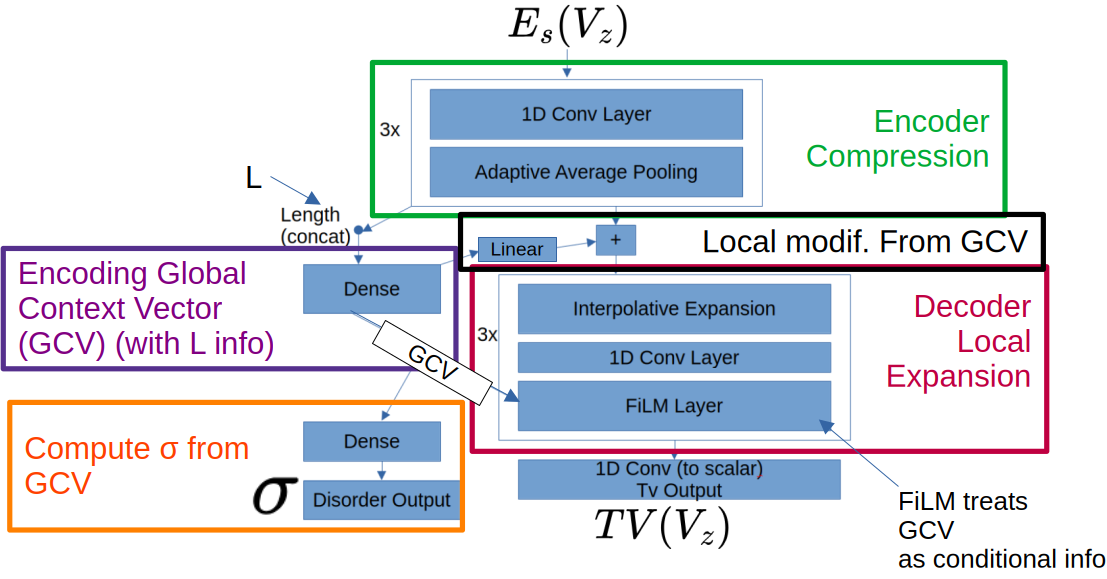}
    \caption{Explanatory diagram of Supervised learning Neural Network. The neural network consists of 5 parts: the $E_s(V_z)$ encoder (in green) that converts the splitting into a smaller embedding space, the global context vector (GCV) (in purple) that converts the embedding along with the length into a small context vector containing global information able to be used as a conditional, the local modification from GCV (in black) that takes the GCV and modifies the embedding space in the locally relevant manner, the decoder (in red) which expands and maps the embedding along with the GCV conditional into the $\text{TV}(V_z)$ output, and finally the small dense neural network in (orange) that converts the GCV to a global disorder magnitude.}
    \label{fig:explantorynn}
\end{figure*}

\section{Neural Network Architectures}
There are two separate neural networks used in this paper. The first is employed for the unsupervised learning task and consists of a standard autoencoder, shown in Fig.~\ref{fig:fig1}. The second is used for the supervised learning task and is a more novel architecture that incorporates global context vectors into an encoder-decoder domain conversion setup, as shown in Fig.~\ref{fig:fig4}(a).
All neural networks are implemented in PyTorch, and scikit-learn. Code is available from the authors upon reasonable request.

\label{sec:nnarchitecture}
\subsection{Unsupervised autoencoder network}
The auto-encoder neural network (shown in Fig.~\ref{fig:fig1}) works by taking in input in the form of two-row array \begin{equation}
    X_{in}=\begin{bmatrix}
        E_s(V_z^{(1)})\ldots E_s(V_z^{(n)}) \\\text{TV}(V_z^{(1)})\ldots \text{TV}(V_z^{(n)})
    \end{bmatrix}
\end{equation} 
The encoder consists of three 1D convolutional layers with $[32, 64, 128]$ filters, respectively. These layers convolve over the $V_z$ axis, where the rows $E_s(V_z)$ and $\text{TV}(V_z)$ are treated as separate input channels. Each convolutional layer uses a stride of 2, halving the size of the $V_z$ axis at each step. The encoder concludes by applying a linear layer that maps to a bottleneck latent space of size 15. This dimensionality was selected because 15 components capture approximately 95\% of the explained variance in a linear PCA analysis \cite{hotelling1933analysis}. We find that our results are qualitatively robust for latent vector sizes in the range $[5, 20]$, although smaller latent dimensions lead to less well-defined phase boundaries. This compressed latent space vector is used by the clustering algorithm to perform unsupervised classification. The decoder is nearly identical to the encoder, but operates in reverse, using transposed convolutional layers to progressively upscale (with stride 2) the representation until the output vector $X_{out}$ matches the original input size $X_{in}$. The training process consists of optimizing the network to achieve $X_{out} = X_{in}$, that is, to learn a compression into a latent space vector $\vec{L}$ from which the original input can be reconstructed through the decoder. The autoencoder network is intentionally simple in order to allow the compression to be natural for the data, and thus to prevent overfitting. The autoencoder is also more stable and less parameter-dependent than alternative kernel methods \cite{hofmann2008kernel}, as it tends to work by naturally finding an encoding of the data in an optimal manner inherent to the training process.

\subsection{Supervised network}
The supervised learning neural network, shown in Fig.~\ref{fig:explantorynn}, is more novel than the unsupervised architecture. The network input consists of a vector $[E_s(V_z^{(1)}) \ldots E_s(V_z^{(n)})]$ of length 512, with $V_z \in [0, 2] \text{ meV}$, while the output consists of an identically sized vector $[\text{TV}(V_z^{(1)}) \ldots \text{TV}(V_z^{(n)})]$, along with a single scalar component representing the disorder magnitude $\sigma$. The network is composed of five distinct parts, as outlined in the figure. The encoder consists of three blocks of 1D convolutional layers, each followed by adaptive average pooling that halves the size of the $V_z$ axis. 

The encoder convolutional layers have 128 filters. The encoder compresses the input data and, when combined with the pooling operations, allows different data $V_z$ scales to be assessed. The network's global context vector encoding, shown in purple, consists of taking the full output of the encoder, concatenating it with the wire length, and feeding this combined representation into a dense layer with 64 neurons to form a small global context vector (GCV) that is used as a conditional input to the decoder. This GCV is also independently fed into another dense layer with 32 neurons, from which the scalar disorder magnitude is computed and output. The GCV is additionally passed through a linear layer and then additively combined with the encoder output to form the input, shown in black, to the decoder. This design allows locally relevant information from the GCV to be incorporated into the input to the decoder, and performance is worse without this inclusion. The decoder, shown in red, consists of three blocks, with each block comprising an interpolative expansion that doubles the size of the $V_z$ axis, followed by a 1D convolutional layer and a FiLM layer \cite{perez2018film} that incorporates the GCV as conditional information into the convolutional process. The FiLM layer operates by applying channel-wise scaling and shifting to each convolutional output based on the GCV, thereby allowing the convolutional network to effectively utilize the global context information. These convolutional layers have 128 filters. The network concludes with a single convolutional layer that combines the multiple decoder channels into a single output vector $\text{TV}(V_z)$.
 
\section{Clustering results using Majorana splitting energy only}
\label{sec:onlyes}
\begin{figure}[ht]
    \centering
    \includegraphics[width=\linewidth]{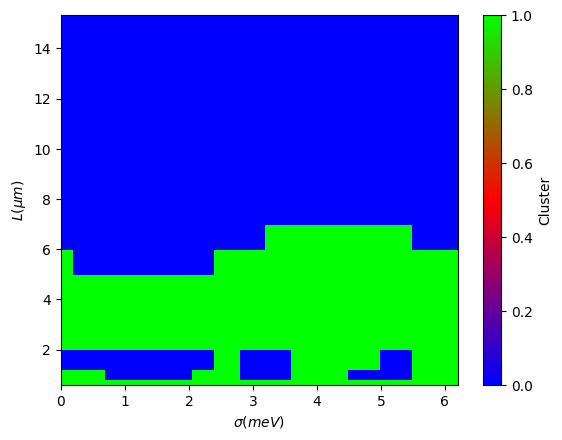}
    \caption{Unsupervised phase diagram for input data lacking $\text{TV}$ in the disorder-length ($\sigma$-$L$) parameter space, obtained by mapping latent space clusters back to physical parameters. Two-cluster classification reveals trivial clustering only between short and long systems, with some noise yielding to unclean boundaries. Forcing more than 2 clusters yields unpredictable results.}
    \label{fig:NoTV}
\end{figure}

The cluster result in the main text uses both the topological visibility $\text{TV}(V_z)$ and the Majorana splitting energy $E_s(V_z)$ as input to the autoencoder. We find that the inclusion of $\text{TV}(V_z)$ is crucial for a meaningful clustering result. If we only use $E_s(V_z)$ as input to the autoencoder, the clustering result becomes trivial, as shown in Fig.~\ref{fig:NoTV}. 

\begin{figure}[ht]
    \centering
    \includegraphics[width=\linewidth]{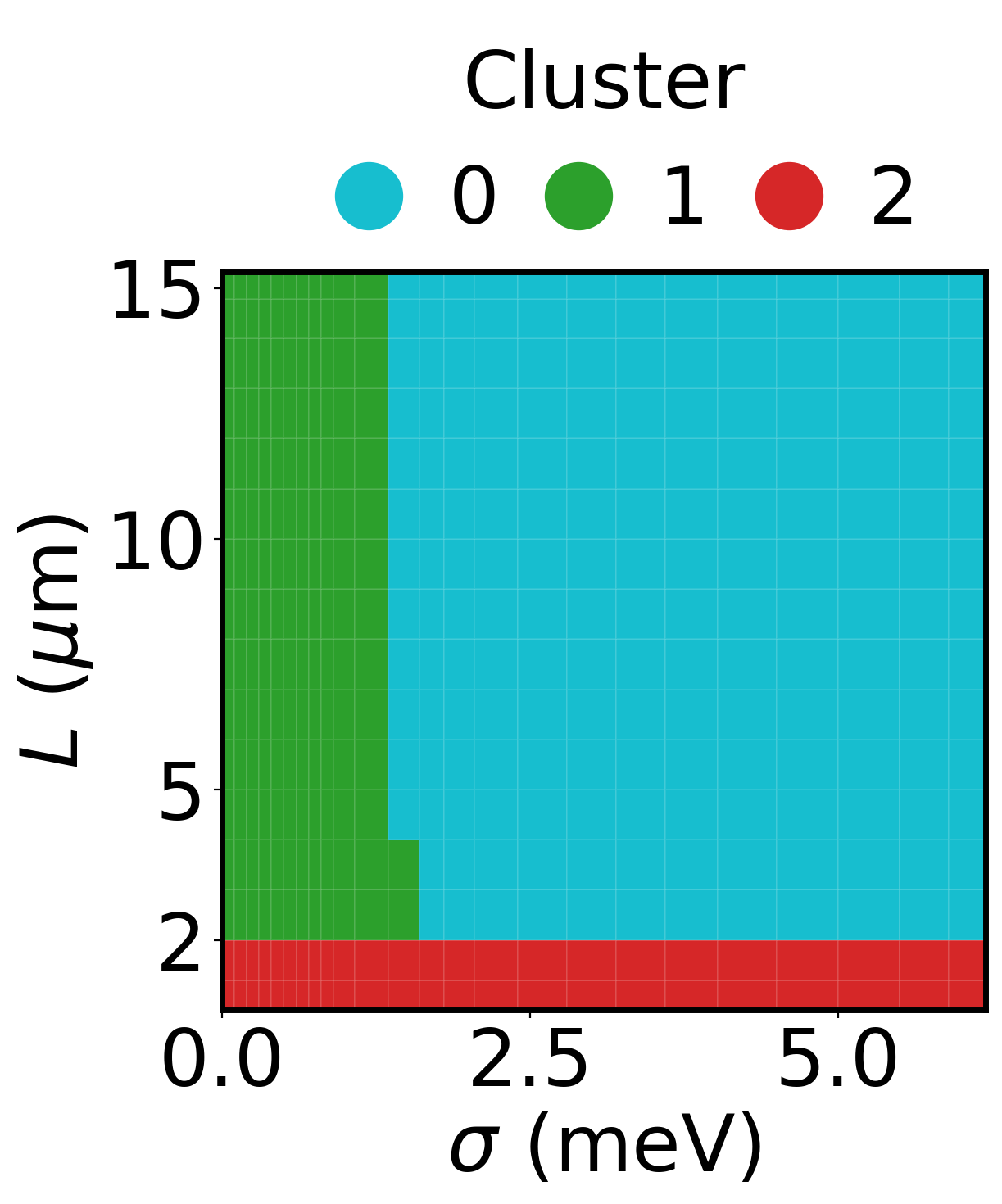}
    \caption{Unsupervised phase diagrams in disorder-length ($\sigma$-$L$) space, obtained by mapping latent-space clusters back to physical parameters using $E_s$ and $L$ as inputs (but only rewarding $E_s$ reconstruction). (a) Three-cluster classification separates weak-disorder, strong-disorder, and short-length regimes; each point is labeled by the majority cluster over 100 disorder realizations. See Fig.~\ref{fig:figUSLNOTV} for four-cluster results and silhouette scores. }
    \label{fig:USLNoTVCluster3}
\end{figure}

The USL can be made to work, though less effectively than the TV-inclusive case, by modifying the auto-encoder scheme. In this approach, $E_s(V_z)$ is used as one input channel, while a second channel contains $L$. During training the auto-encoder is rewarded only for reconstructing $E_s(V_z)$. The rationale is to prevent the network from forming clusters based primarily on length. With this setup, one obtains clustering phases that are approximately similar to those of the TV-inclusive case while still optimizing the silhouette score (see Fig.~\ref{fig:figUSLNOTV} and Fig.~\ref{fig:USLNoTVCluster3}).
In this case, the result is less stable than in the TV-inclusive case, since it is necessary to remove very short wires, with $L<1~\mu $m, from the dataset; otherwise, the algorithm forms superfluous short-wire clusters. This is not particularly surprising, since $L$ is part of the input and the splitting depends strongly on it. The TV-inclusive procedure outlined in the main text is likely more reliable, but this alternative scheme allows direct experimental data to be used without any supervised learning.

\section{Accuracy of the supervised learning}
\begin{figure}[ht]
    \centering
    \includegraphics[width=\linewidth]{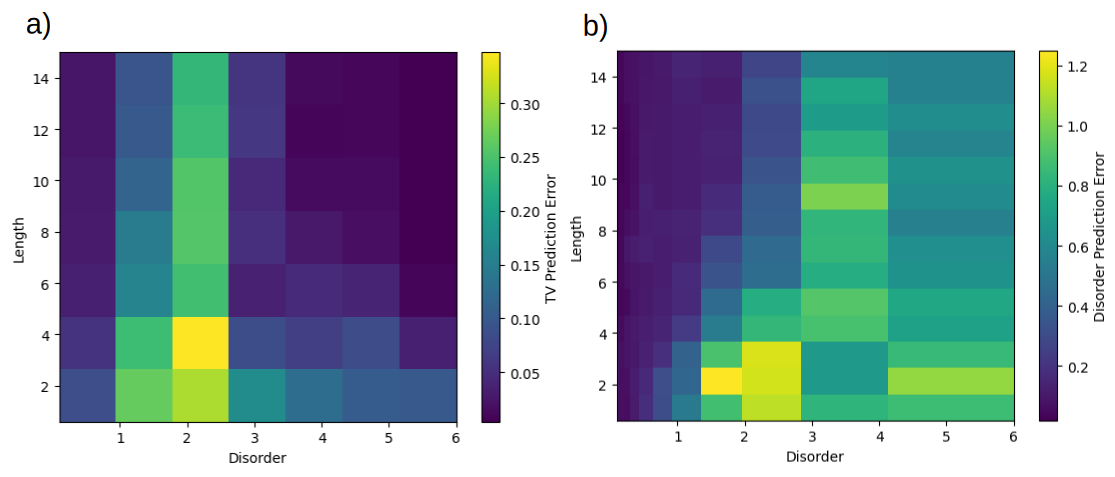}
\caption{Prediction error heatmaps in the ($\sigma$, $L$) parameter space for the full disorder regime $\sigma\in [0,6]$ meV: (a) $\text{TV}$ prediction error, (b) disorder magnitude prediction error. Color indicates the magnitude of prediction error (lighter/yellow = higher error). Interestingly, errors appear larger near the phase transition resulting from the unsupervised learning. }
    \label{fig:accuracy_strong}
\end{figure}

\begin{figure}[ht]
    \centering
    \includegraphics[width=\linewidth]{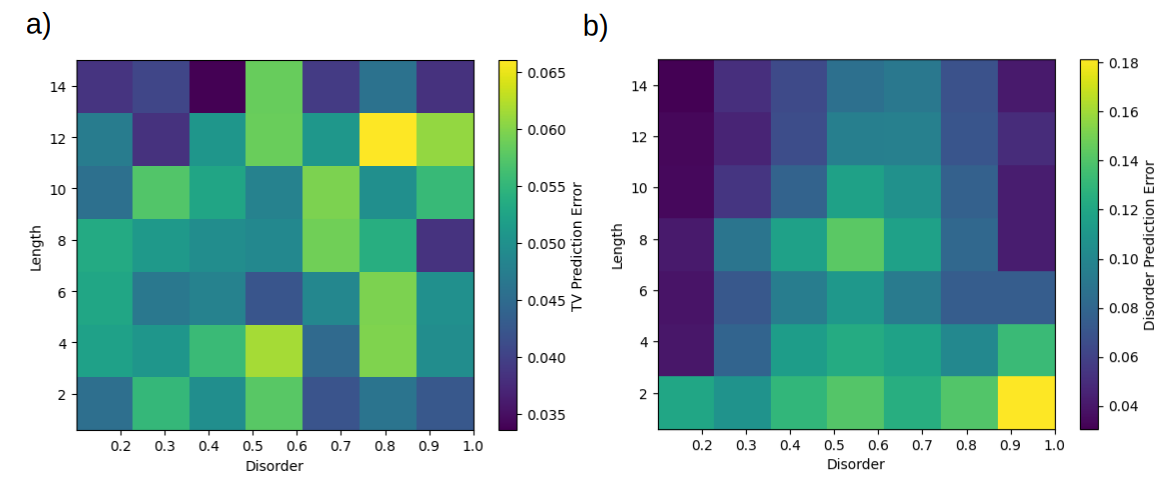}
\caption{Prediction error heatmaps in the ($\sigma$, $L$) parameter space for weak disorder regime $\sigma\in [0,1]$ meV: (a) $\text{TV}$ prediction error, (b) disorder magnitude prediction error. Color indicates the magnitude of prediction error (lighter/yellow = higher error). Errors are generally larger in the high disorder, short wire length regime. }
    \label{fig:accuracy_weak}
\end{figure}

The prediction accuracy of the supervised model as a function of the disorder magnitude $\sigma$ and wire length $L$, corresponding to Fig.~\ref{fig:fig4} for the strong-disorder regime $\sigma \in [0,6]$ meV and the weak-disorder regime $\sigma \in [0,1]$ meV, is shown in Fig.~\ref{fig:accuracy_strong} and Fig.~\ref{fig:accuracy_weak}, respectively.

\section{Silhouette Score}\label{sec:silhouette}
Let $\{x_i\}_{i=1}^N$ be points (e.g., latent vectors) partitioned by $k$-means into $k\ge 2$ clusters $\{\mathcal C_1,\dots,\mathcal C_K\}$ with index map $k(i)$ for point $x_i$. Let $d(\cdot,\cdot)$ denote the distance in latent space (Euclidean unless stated otherwise). For a point $x_i$ define the mean intra-cluster distance
\begin{multline}
  a(i) = \frac{1}{\lvert \mathcal C_{k(i)}\rvert-1} \sum_{\substack{x_j\in \mathcal C_{k(i)}\\ j\ne i}} d(x_i,x_j)\, ,\quad \\\text{(use }a(i)=0\text{ if }\lvert \mathcal C_{k(i)}\rvert=1)\, .
\end{multline}
and the smallest mean distance to another cluster
\begin{equation}
  b(i) = \min_{\ell\ne k(i)}\; \frac{1}{\lvert \mathcal C_{\ell}\rvert} \sum_{x_j\in \mathcal C_{\ell}} d(x_i,x_j)\, .
\end{equation}
The silhouette of $x_i$ and the overall silhouette score are
\begin{equation}
  s(i) = \frac{b(i)-a(i)}{\max\{a(i),\, b(i)\}}\in[-1,1] \quad \text{and} \quad S= \frac{1}{N}\sum_{i=1}^N s(i)\, .
\end{equation}
Interpretation: $s(i)\approx 1$ means $x_i$ is well matched to its cluster; $s(i)\approx 0$ indicates boundary/overlap; $s(i)<0$ suggests misassignment. Larger $S$ indicates better-defined clusters (often $S\gtrsim 0.5$ is considered good, but this is data dependent).

\section{$k$-means}\label{sec:kmeans}
The $k$-means algorithm is employed to cluster the latent space representations obtained from the autoencoder. The goal is to group similar disorder realizations together based on their latent space features.

The $k$-means algorithm~\cite{ikotun2023kmeans} works by initializing $k$ centroids in the latent space, and iteratively refining their positions based on the data points assigned to each cluster. The assignment step involves computing the distance between each data point and the centroids, while the update step recalculates the centroids as the mean of the assigned points. 

To determine the optimal number of clusters $k$, we utilize the silhouette score discussed above, which measures how similar an object is to its own cluster compared to other clusters. A higher silhouette score indicates better-defined clusters. This process should ideally yield well-separated clusters of points within the latent space.

\section{Semi-Supervised results}
\label{sec:semisupervised}
\begin{figure}[htbp]
    \centering
    \includegraphics[width=1.0\linewidth]{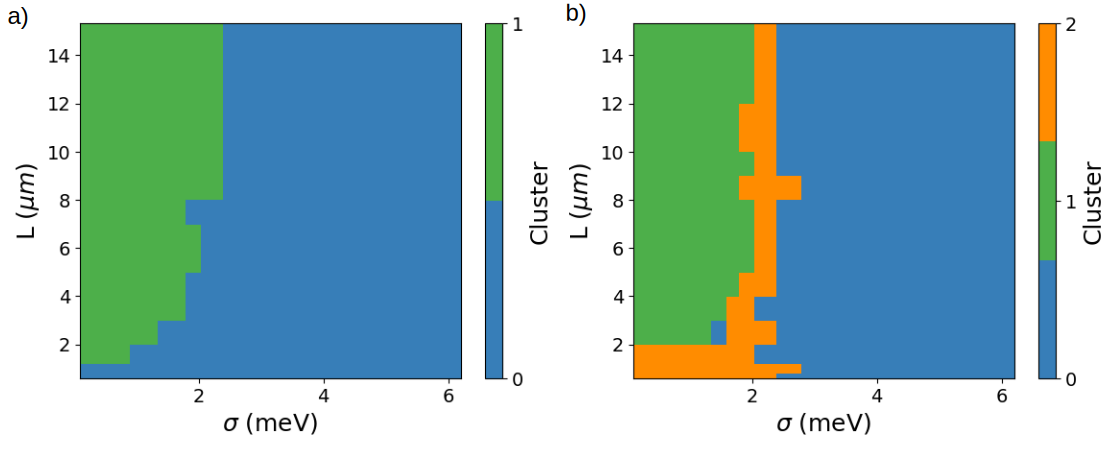}
    \caption{ 
    Semi-supervised phase diagrams in the disorder-length ($\sigma$-$L$) parameter space, obtained by mapping latent space clusters back to physical parameters. The input consisted entirely of $E_s$, with $\text{TV}$ being predicted by the supervised network. Both $E_s$ and the predicted $\text{TV}$ were then fed into the unsupervised process. Results are in good qualitative agreement with those obtained using the exact $\text{TV}$ values.
    (a) Two-cluster classification reveals a phase boundary between weak-disorder (Cluster 1, green) and strong-disorder (Cluster 0, blue) regimes.
    (b) Three-cluster classification identifies an additional intermediate regime (Cluster 2, orange) at the crossover.
    Each point is labeled by the majority cluster among disorder realizations.
    }
    \label{fig:figfullcycle}
\end{figure}

Experimentally, $\text{TV}$ is generally unknown and thus cannot be directly used to classify devices into specific topological phases. In this section, we present hybrid supervised–unsupervised (semi-supervised) results that rely solely on $E_s(V_z)$ as input, combined with a trained supervised network to classify devices. The overall process is: $$E_s \;\xrightarrow{\text{Supervised}} \text{TV}^{\text{Pred}}
\;\;\xrightarrow{\;\;\text{Unsupervised}(E_s,\;\text{TV}^{\text{pred}})\;\;}
\text{Phase}$$
Specifically, we utilize our supervised neural network to predict $\text{TV}(V_z)$ from $E_s(V_z)$ and then feed both into our unsupervised process to output the topological phase. We find that the results are in good qualitative agreement with those obtained using the exact $\text{TV}(V_z)$ as input, reproducing the same phase boundaries but with less sharply defined transitions, as expected from using predicted values (See Fig. \ref{fig:figfullcycle}). The only qualitative difference is in the very short length 2-cluster scenario, where the topology is ill-defined in the first place.
\begin{figure}[htbp]
    \centering
    \includegraphics[width=1.0\linewidth]{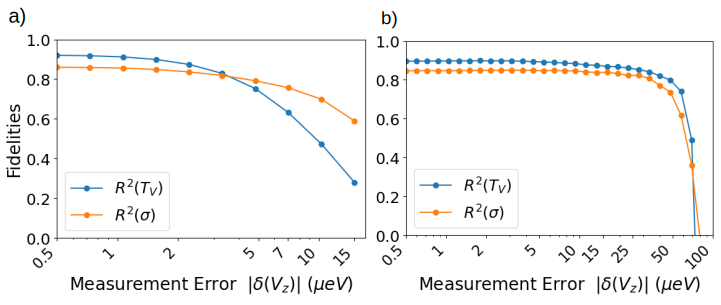}
    \caption{Average $R^2(\text{TV})$ vs. $|\delta|$, where $|\delta|$ is the standard deviation of additive error $E_s'(V_z)=E_s(V_z)+\delta(V_z)$ for (a) the standard neural network (b) the neural network with error added during training. The neural network shows robust resistance to measurement errors within $E_s(V_z)$ up to (a) $5\mu eV$ and (b) $50 \mu eV$.}
    \label{fig:figRobustness}
\end{figure}

\end{document}